# GEWUM: General Exploration Workflow for the Utopia of Materials: A Unified Platform for Automated Structure Generation, Selection, and Validation

Jiexi Song [a, #], Aixian She [b, d,#], Changpeng Song [b,d], Diwei Shi [c,*], Fengyuan Xuan [a,*] and Chongde Cao [b,d,*]

a Suzhou Laboratory, Suzhou 215123, China
b Research and Development Institute of Northwestern Polytechnical University in Shenzhen, Shenzhen 518057, China
c School of Naval Architecture and Maritime, Zhejiang Ocean University, Zhoushan 316022, China
d School of Physical Science and Technology, Northwestern Polytechnical University, Xian 710072, China
#: Jiexi Song and Aixian She contributed equally
Author to whom correspondence should be addressed:
shidiwei@zjou.edu.cn; xuanfy@szlab.ac.cn; caocd@nwpu.edu.cn

## ABSTRACT

The fragmented landscape of existing computational tools often hinders the seamless integration of large-scale structure prediction with rigorous stability validation. To address this, we present GEWUM (General Exploration Workflow for the Utopia of Materials), an open-source platform that integrates the Selective Random Structure Search (SRSS) strategy with universal Machine Learning Interatomic Potentials (uMLIPs) to automate and accelerate materials discovery. With native support for SLURM-based HPC clusters, GEWUM unifies the entire workflow from structure generation and diversity-preserving selection to thermodynamic/dynamic stability assessments and property calculations. The platform further incorporates built-in visualization tools, including Sankey diagrams for space-group evolution, violin plots for energy distributions, and t-SNE/UMAP embeddings for structural diversity, enabling intuitive interpretation of screening results. We demonstrate GEWUM through three case studies: low-energy polymorph prediction in Al-Sc-N, identification of a *P*-62*c* phase of $U_3Si_5$, and high-pressure structure prediction of $ThH_{10}$ at 150 GPa. Benchmark tests confirm reasonable agreement in thermophysical property predictions.

# I. INTRODUCTION

The quest for novel materials with tailored properties has long been driven by the need to address global challenges in energy, electronics, and sustainability [1-4]. In recent years, the landscape of computational materials science has undergone a paradigm shift with the advent of universal Machine Learning Interatomic Potentials (uMLIPs) [5-9]. Unlike traditional empirical potentials, pre-trained uMLIPs now offer near-density functional theory (DFT) accuracy at a fraction of the computational cost, enabling the rapid exploration of vast chemical spaces [10-12]. Consequently, leveraging these advanced potentials for both crystal structure prediction (CSP) and subsequent property evaluation has emerged as a dominant strategy in modern materials discovery [13-19].

However, despite the maturity of uMLIP algorithms and the availability of various CSP codes capable of handling diverse dimensionalities (from 0D molecules to 3D crystals), a critical gap remains in the software infrastructure required to fully exploit this potential [20-25]. Currently, researchers face a fragmented ecosystem: while individual tools may excel at structure generation or single-point property calculations, there is a distinct lack of unified, user-friendly platforms that seamlessly integrate the entire workflow [26-29]. Notably, efforts such as AFLOW and the online tools provided by MaterialsAtlas.org [30-33] can be combined into pipelines that couple structure generation with property evaluation, representing valuable contributions toward workflow integration. That said, these platforms primarily focus on high-throughput screening or database-driven workflows, and do not yet provide a unified, uMLIP-native framework for crystal structure prediction with built-in HPC orchestration. Specifically, existing approaches often require users to manually stitch together disparate scripts for structure sampling, diversity selection, and rigorous stability validation (e.g., convex hull analysis and phonon calculations). More critically, most frameworks lack native integration with high-performance computing (HPC) workload managers like SLURM. This absence of built-in parallelization capabilities makes it exceedingly difficult to scale uMLIP-driven workflows from testing a few candidates to screening thousands of structures across complex chemical spaces, thereby stifling the efficiency gains promised by machine learning.

To bridge this gap and empower the materials community with a tool that is both easy to adopt and highly efficient, we present a unified platform, GEWUM (General Exploration Workflow for the Utopia of Materials). GEWUM is designed not merely as a structure predictor, but as a comprehensive, end-to-end software platform engineered to maximize the utility of uMLIPs in large-scale scientific research. By embedding the recently developed Selective Random Structure Search (SRSS) strategy [34] within a robust engineering framework, GEWUM eliminates the need for manual script management and provides a streamlined pathway from initial hypothesis to validated material candidate.

The core strength of GEWUM lies in its holistic integration, user-centric design, and scalability. First, it unifies the entire discovery pipeline: the software automatically handles dimensionality-agnostic structure generation, adaptive diversity selection, and crucially thermodynamic and dynamical stability assessments. This "predict-and-verify" architecture ensures that users receive physically realizable candidates without the burden of disconnected post-processing steps. Second, and most significantly, GEWUM features seamless integration with SLURM clusters through a simplified configuration mechanism. Users can enable massive parallelism across the entire workflow, from structure relaxation to property calculation, simply by modifying a single template file. This design abstracts away the complexities of job submission scripts, resource allocation, and dependency management, allowing researchers to effortlessly harness hundreds or thousands of CPU cores without any specialized HPC expertise. Finally, GEWUM prioritizes accessibility through a modular command-line interface and pre-configured templates for complex properties (e.g., elastic constants, thermal conductivity, and quasi-harmonic approximations), enabling scientists to focus on physical insights rather than computational intricacies.

In this work, we detail the architecture and capabilities of GEWUM, demonstrating its effectiveness through the discovery of previously unknown polymorphs in diverse systems. By combining the unbiased exploration power of SRSS, the speed and accuracy of uMLIPs, and the effortless scalability of its SLURM integration, GEWUM establishes a new standard for efficient, comprehensive, and user-friendly materials discovery, effectively unlocking the full potential of machine learning potentials for the broader scientific community.

# II. Software Design and Implementation

## 2.1 Software Architecture

### 2.1.1 Design Philosophy

GEWUM adopts a modular design philosophy that decomposes complex materials computational workflows into independent yet composable functional modules. The architectural design adheres to three core principles: unified interface, flexible deployment, and template-driven configuration. The unified interface principle ensures that all workflows are accessible through a consistent command-line paradigm (*gewum <command> --mode <mode>*), significantly reducing the learning curve for users and enabling seamless navigation across different computational modules. The flexible deployment principle empowers users with fine-grained control over computational scripts through a file-copying mechanism, allowing local modifications, debugging, and customization without altering the core software installation. The template-driven principle addresses the challenge of cross-platform computational environment adaptation by introducing a hierarchical SLURM configuration system, where deployment-specific parameters are automatically injected into computational scripts through placeholder substitution.

### 2.1.2 System Architecture

The GEWUM platform is architected as a four-tier hierarchical system that cleanly separates user interaction, workflow orchestration, script management and computational execution. The workflow for the client-side usage is illustrated in . 1(a). The Command Line Interface constitutes the uppermost layer and primary interface for user interaction. It exposes seven principal workflow commands (RD, PT, ELA, QHA, TC, MD, DB) through a consistent syntactic paradigm. Beneath this lies the Workflow Command layer, which implements an object-oriented command dispatching system centered around the base class BaseWorkflowCommand. This layer encapsulates common functionalities such as argument parsing, file repository management, and SLURM template processing, while concrete implementations (RDCommand, PTCommand, etc.) handle workflow-specific logic. The Script Repository layer constitutes the functional backbone of the system, organizing computational scripts into workflow-specific directories (RDworkflows, PTworkflows, etc.) and shared functional modules under the common directory. The shared modules encompass critical cross-cutting concerns including structure relaxation, multi-dimensional structure selection, phonon calculation, energy hull analysis, and post-processing symmetry analysis. At the foundation, the Computational Backend layer integrates external scientific computing libraries and machine learning frameworks, with built-in pretrained universal machine learning force field calculations, PyXtal enabling symmetry-constrained crystal generation [35], Phonopy and Phono3py handling lattice dynamics and thermal transport [36], and ASE serving as the unified atomic simulation environment interface [37]. The home page snapshot was displayed in Fig. 1(b).

### 2.1.3 Core Components

The architectural integrity of GEWUM relies upon several core components that collectively ensure system reliability, extensibility, and user-friendliness. The Command Dispatcher, implemented in *main.py*, serves as the central nervous system of the platform, dynamically loading workflow modules through introspection, orchestrating parameter validation and parsing, and providing unified error handling and logging facilities. The BaseWorkflowCommand class establishes interface for all workflow implementations, defining standard methods for script discovery, template rendering, and file deployment, while abstracting the complexities of SLURM configuration integration. The Template Utilities module (*template_utils.py*) implements configuration resolution logic, searching for user-defined settings across multiple hierarchical levels (working directory, installation directory, and built-in defaults), and automatically generating SLURM header blocks with appropriate resource allocation directives and environment setup commands. The Configuration Manager (*config.py*) centralizes path management and maintains the definitive mapping between workflow commands and their associated script repositories, ensuring consistent file resolution across diverse deployment scenarios. Together, these components form a cohesive architectural foundation that balances abstraction with flexibility, enabling rapid development of new workflow modules while preserving the operational consistency that characterizes the GEWUM user experience.

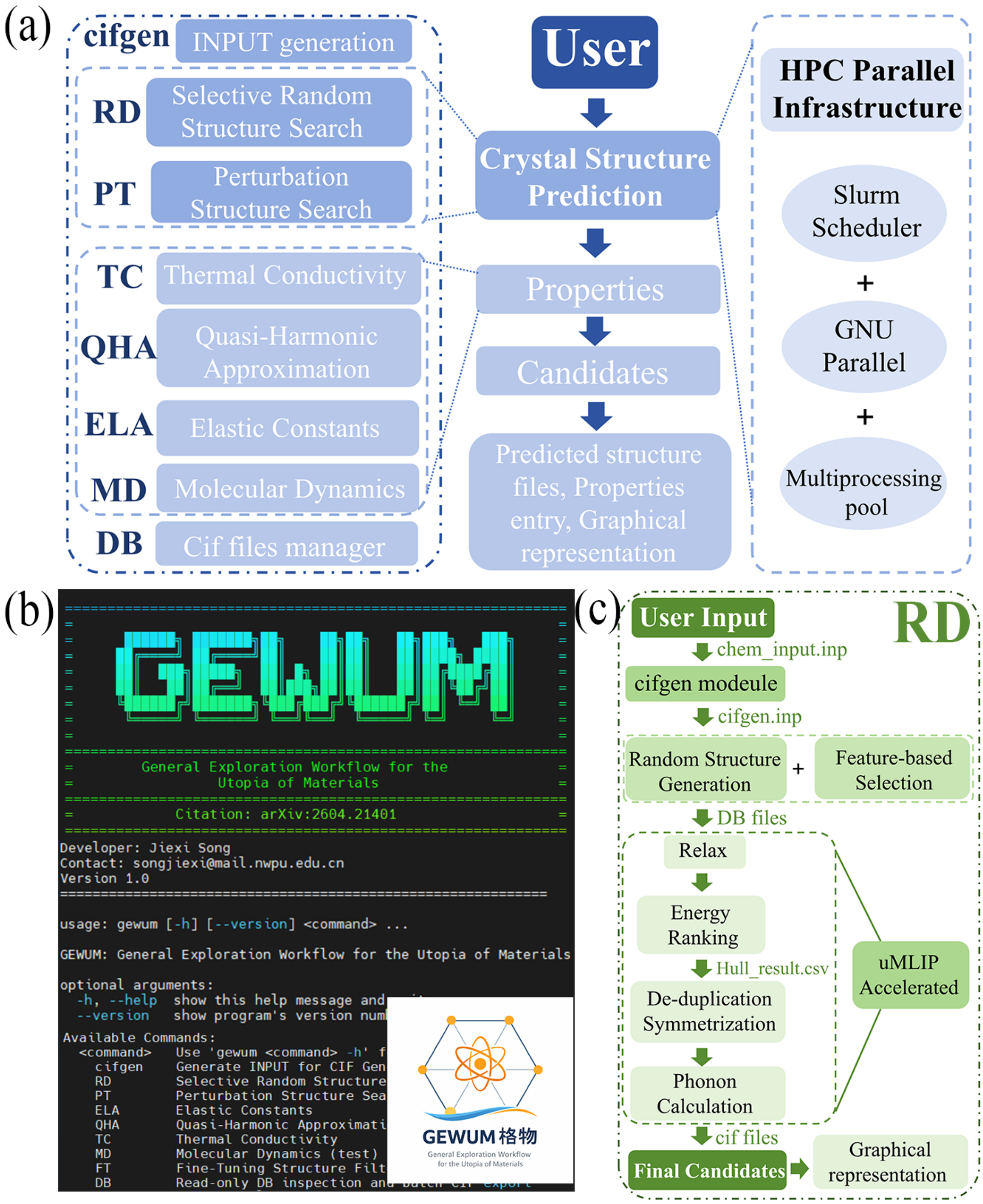

**Fig 1.** (a) The system architecture of GEWUM. (b) Home page snapshot; the inset shows the GEWUM logo. (c) The workflow of the RD module.

## 2.3. Core Functionality

### 2.3.1 Random Design Workflow (RD)

A primary capability of GEWUM is the implementation of the SRSS framework for crystal structure generation and screening, see Fig. 1(c). This method can explore configurational space through symmetry-constrained random generation encompassing 230 space groups, 80 layer groups, 75 rod groups, and 32 point groups, followed by feature-based diversity selection using K-means [38] clustering or HDBSCAN [39] density-based clustering. Alternatively, one can bypass the selection step and directly optimize all structures. The selected candidates undergo high-throughput geometric relaxation using uMLIPs. The SRSS framework employs a dual-filtering stability assessment mechanism based on thermodynamic stability via convex hull energy calculation against existing structures in the Materials Project database [40] and dynamic stability verification through phonon dispersion analysis with automatic elimination of structures exhibiting imaginary frequencies. For 2D materials, layer group symmetry classification with automatic segregation into centrosymmetric and non-centrosymmetric categories is provided, while three-dimensional structures undergo space group symmetrization with optional chiral identification. Consequently, GEWUM can identify stable and metastable structures across all periodic dimensionalities, from bulk crystals to low-dimensional nanomaterials, requiring only standard computational resources.

### 2.3.2 Perturbation Workflow (PT)

The Perturbation workflow enables systematic exploration of structural phase space from known crystallographic configurations, supporting diverse materials engineering applications including phase transition studies, doping engineering, and compositional substitution screening. The module implements four distinct structure generation paradigms: supercell generation for studying finite-size effects and constructing defect models through controlled expansion of unit cell dimensions;

element substitution based on user-defined replacement rules specified in YAML configuration files, enabling systematic exploration of isovalent and aliovalent substitutions; structural mutation through controlled random displacement of atomic positions and application of lattice strain perturbations, facilitating escape from local energy minima and exploration of adjacent regions of the potential energy surface; and doping engineering through concentration-controlled random replacement of host atoms with dopant species, automatically generating multiple doping configurations while preserving overall charge neutrality. A critical architectural feature of the PT workflow is its seamless interoperability with the RD workflow's screening output, structures identified through RD screening can be directly channeled into the PT pipeline for high-throughput elemental expansion and structural exploration, thereby maximizing the discovery of diverse chemical configurations. The workflow of PT module can be found in the suplementary materials.

### 2.3.3 Property Calculations

GEWUM integrates a comprehensive suite of property calculation modules (ELA, QHA, TC, MD) that leverage uMLIPs through ASE's unified Calculator interface, enabling efficient characterization of mechanical, thermal, and thermodynamic properties without the computational cost of traditional DFT methods. The Elastic Analysis (ELA) module implements the energy-strain methodology through integration with Vaspkit's deformation analysis tools, calculating complete elastic constant tensors from stress responses to applied strains and deriving bulk modulus, shear modulus, Young's modulus, and Poisson ratio with support for both small-strain and finite-strain calculation modes. The lattice dynamics and thermal transport capabilities are provided by the Thermal Conductivity (TC) module, which combines uMLIP force evaluations with Phonopy and Phono3py to deliver phonon dispersion relations, density of states, and temperature-dependent lattice thermal conductivity via the Boltzmann transport equation in the relaxation time approximation. The Quasi-Harmonic Approximation (QHA) module automates volume-dependent free energy calculations across a series of lattice expansion points, enabling prediction of thermal expansion coefficients and temperature-dependent heat capacities. The Molecular Dynamics (MD) module provides NVT ensemble simulations with Langevin thermostat integration for assessing high-temperature kinetic stability, generating energy-time evolution trajectories and temperature fluctuation statistics for structural degradation analysis. These property calculation modules share a unified computational backend through ASE's Calculator interface, ensuring consistent force evaluation and seamless integration with the SRSS workflow for comprehensive materials characterization. Additionally, the Database (DB) module provides a read-only interface for inspecting and managing structures.db contents, enabling users to query metadata, perform multi-dimensional statistical analysis, and filter structures based on chemical or energetic criteria. It further facilitates data portability by supporting the batch export of specific or filtered CIF files to local directories or ZIP archives without modifying the underlying database.

### 2.4. Parallel Execution and Fault Tolerance

#### 2.4.1 Two-Level Parallelism Architecture

GEWUM employs a hierarchical two-level parallel execution architecture designed to maximize computational throughput across distributed HPC environments. At the outer level, all workflows are natively integrated with the SLURM job scheduler through templated submission scripts, where resource allocation directives are dynamically parameterized via a unified *slurm_config.yaml* configuration file. This design enables seamless deployment on heterogeneous HPC clusters without requiring manual modification of individual workflow scripts. At the inner level, CPU-intensive computational tasks are parallelized across all available cores within a single compute node using either Python's multiprocessing.Pool or GNU Parallel [41-45], with the degree of concurrency automatically inferred from the SLURM_CPUS_PER_TASK environment variable to ensure optimal resource utilization.

### 2.4.2 Workflow-Specific Parallelization Strategies

The platform implements specialized parallelization strategies tailored to the computational characteristics of each workflow stage. Structure generation via *cif_generate.py* exploits parallelism at the symmetry-group level, distributing tasks across all target symmetry groups using *multiprocessing.Pool.map()* to achieve near-linear scaling with available CPU count. Structure relaxation employs a task-queue paradigm where all unprocessed CIF files are enumerated into a flat task list with automatic detection and skipping of previously relaxed structures to support fault-tolerant restart; GNU Parallel then dispatches concurrent relaxation jobs, ensuring full utilization of allocated computational resources. The structure diversity selection module processes multiple symmetry-group subdirectories concurrently using a configurable worker pool, supporting all dimensionalities from 0D to 3D and multiple descriptor-clustering combinations without sacrificing throughput.

### 2.4.3 Fault Tolerance and Configuration Management

To accommodate the scale of high-throughput computational campaigns, all batch scripts incorporate idempotent task filtering mechanisms that detect and skip structures already processed during previous execution attempts. This design enables safe resubmission following node failures or walltime limit violations without incurring redundant computation. Furthermore, all SLURM parameters, including partition selection, CPU allocation, wall time limits, conda environment activation, and module loading sequences are centrally managed within the slurm_config.yaml file and injected into every workflow script at generation time through a template substitution engine. This centralized configuration approach eliminates per-script configuration drift and allows users to adapt the entire GEWUM suite to different HPC environments through modification of a single file.

## 2.5. Key Input Files

GEWUM relies on a concise set of user-facing input files that govern workflow behavior across all modules. The primary configuration file, *slurm_config.yaml*, centralizes all HPC environment parameters including SLURM resource directives, module loading sequences, and conda environment paths, and serves as the single point of adaptation when deploying GEWUM on a new computational cluster. For structure generation, *chem_input.inp* specifies the target chemical compositions, stoichiometric ratios in a lightweight line-based format, from which the intermediate *cifgen.inp* is automatically generated for further structures' generation. The PT utilize two additional

YAML-format [46-49] configuration files: *replacements.yaml* defines element substitution rules for systematic compositional exploration, while *doping.yaml* specifies dopant species, host elements, substitution concentrations, and the number of doping configurations to generate. Together, these files constitute the complete user-facing configuration layer of GEWUM, intentionally minimized to lower the barrier for routine use while retaining sufficient flexibility for advanced customization.

# III. Examples

### 3.1 Case 1：Structure Prediction for the Al-Sc-N System

The wurtzite Al-Sc-N system has attracted considerable interest owing to its prospective applications in piezoelectric, ferroelectric, and functional thin-film materials [50-53]. To assess the capability of GEWUM for structure prediction in complex multicomponent nitride systems, we select $(Al_{13}ScN_{16})x$ (x=1, 2) as the model system. The overall computational workflow is illustrated in Fig. 2(a).

Step1: Candidate Structure Prediction

Within GEWUM, the user is required only to specify the target chemical composition and dimensionality, after which the software automatically constructs candidate crystal structures consistent with the compositional constraints. For the target compositions, the input *chemi_input* format is

*Al*
*=13*
*Sc*
*=3*
*N*
*=16*

And the Maximum atomic number of a unit cell is set to 64 atoms.

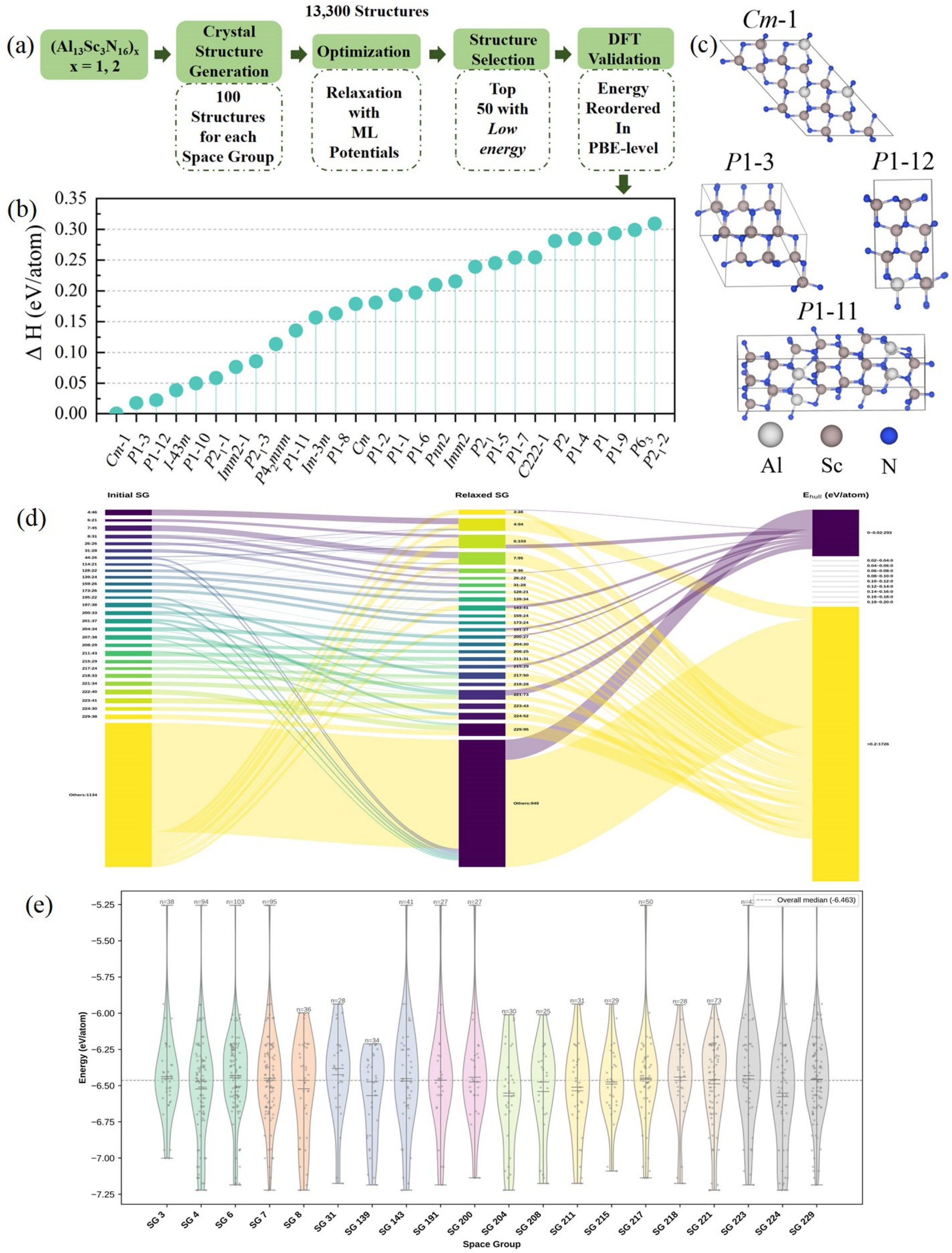

**Fig 2.** (a) The screening workflow of $(Al_{13}ScN_{16})x$ (x=1, 2). (b) The energy difference of predicted structures at the DFT-PBE level. (c) Schematic diagrams of some prominent metastable crystals. (d) The Sankey diagram of changes in the number of structures and the convex-hull energies. (e) Distribution of space groups and energies of structures after optimization.

In principle, these target compositions allow the generation of 229 (space groups) ×100 (trial attempts) ×2 (compositions) =45,800 random structures. However, because a maximum limit of 64 atoms was imposed on the generated structures, and because certain space groups cannot yield trial structures under the simultaneous constraints of target stoichiometry, atom count, and Wyckoff-site restrictions, the actual number of generated candidate structures was 13,200. Use the following command:

*gewum RD –mode cifgen*

Then modify the relevant parameters in *cifge.sh* and submit them to the Slurm cluster for high-throughput generation of crystal structures.

Step2: Rapid Relaxation with a Machine-Learning Potential

GEWUM performs high-throughput geometric optimization by means of the pretrained uMLIP machine-learning potential. Use the following command:

*gewum RD –mode relax*

Then modify the relevant parameters in *relax_umlip.sh* and submit to the cluster for high-throughput optimization of crystal structures.

Step3：Diversity Preservation and Low-Energy Structure Screening

Upon completion of structural relaxation, the convex-hull energy ($E_{hull}$) of each configuration is further evaluated, and the structures are ranked according to $E_{hull}$. The 50 lowest-energy candidate structures are then retained for the subsequent stage of calculation.

Use the following command:

*gewum RD –mode post*

and *gewum RD –mode Ehull*

Step4:DFT Refinement and Final Energy Reordering

The 50 low-energy candidate structures selected in the preceding step are further examined first-principles calculations at the PBE level. Following full structural relaxation, the relative energies of all retained configurations are recalculated and reordered. As shown in Fig. 2(b), the relative enthalpy distribution of the final candidates lies within approximately 0-0.31 eV/atom. Among them, the Cm-1 structure emerges as the ground-state configuration. Meanwhile, several other structures, including *P*1-3, *P*1-12, and *P*1-11, are likewise situated within the low-energy region, with schematic representations of their crystal structures displayed in Fig. 2(c).

Figs. 2(d) and 2(e) further demonstrate the statistical visualization capabilities integrated into the SRSS workflow of GEWUM. Fig. 2(d) presents a Sankey diagram showing the evolution of the generated structures from their initial space groups to the space groups obtained after structural relaxation, and further links the relaxed structures to their corresponding convex-hull energy Ehull distributions. This visualization provides an intuitive representation of possible symmetry evolution during structural optimization and the energetic stability of different structures. Fig. 2(e) employs violin plots to illustrate the energy distributions of the relaxed structures grouped by space group, enabling comparison of the energy ranges, distribution widths, and relative concentrations associated with different structural symmetries. Together, these visualization tools allow users to directly inspect statistical information on space-group evolution, structural populations, and energy distributions in the structure-prediction results, thereby facilitating the identification of representative low-energy structures and potentially stable structural types.

These results indicate that the Al-Sc-N system possesses a remarkably rich configurational landscape. This feature is of considerable significance for elucidating phase competition in Al-Sc-N-based functional materials, identifying plausible synthesis pathways, and clarifying the subsequent relationship between structure and properties.

### 3.2 Case 2：Searching for the Unknown Allotropes of $U_3Si_5$

For a long time, the crystal structure of $U_3Si_5$ has been widely recognized as a hexagonal structure of the β-$ThSi_2$ or $AlB_2$-type (space group P6/mmm) [54], that is, this structure can be described as a defective configuration featuring a single-site silicon vacancy within a 3×1×1 supercell of $USi_2$, as shown in Fig. 3(e). However, recent experiments have confirmed that $U_3Si_5$ actually exhibits two thermodynamically stable phases, indicating that, in addition to the $AlB_2$-type structure, an unknown crystal configuration also exists [55].

Figs. 3(a-d) further illustrate the low-dimensional visualization of the structure selection and optimization processes in GEWUM. Fig. 3(a) compares the distributions of the initially generated, selected, and relaxed structures in the low-dimensional t-SNE space. The initially generated structures exhibit a broad and relatively uniform distribution in the feature space, while the selected structures still provide good coverage of the original structural space. This indicates that the selection process substantially reduces the number of candidate structures without significantly compromising the diversity of structural sampling. Figs. 3(b-d) show the density distributions of the initially generated, selected, and relaxed structures in the t-SNE

space, respectively. The overall distribution characteristics of the initially generated and selected structures remain similar, further demonstrating that the selection strategy preserves representative sampling of the structural space while reducing the computational cost of subsequent calculations. In contrast, after structural optimization, both the overall distribution and local clustering characteristics of the structures in the low-dimensional feature space change markedly, indicating that structural relaxation leads to substantial rearrangements of atomic configurations and local structural environments. Therefore, adequate structural optimization is an essential step in the structure-prediction workflow and plays an important role in obtaining physically meaningful candidate structures and reliably evaluating their subsequent stability.

Based on the RD module of the GEWUM software, this study successfully predicted the *P*-62*c*-$U_3Si_5$ structure, as shown in Fig. 3(f). This structure belongs to the hexagonal crystal system: U atoms are arranged periodically in layers along the c-axis, and are uniformly embedded within a continuous covalent framework formed by Si atoms in the ab plane. The structure is stabilized by weak U-Si interactions, reflecting the coexistence of metallic characteristics of U and covalent properties of Si. In the ab plane, Si atoms form a distorted 12-membered ring network via strong covalent bonds; this covalent ring is the key source of the material's mechanical strength, thermal stability, and excellent irradiation resistance.

Table 1 shows that the lattice parameter c of *P*-62*c*-$U_3Si_5$ (4.258 Å) differs significantly from the theoretically calculated value for $AlB_2$-type $U_3Si_5$ (3.947 Å). To account for the strong correlation effects of 5f electrons in U-based systems, we performed further structural refinement and optimization of both structures using the DFT+U method (with an effective correlation energy $U_{eff}$= 4 eV). The results indicate that after correction with $U_{eff}$, the lattice parameters of both structures show markedly improved agreement with the experimentally measured values. However, the lattice parameter of *P-62c*-$U_3Si_5$ still exhibits a non-negligible systematic deviation from the experimentally determined value for $AlB_2$-type $U_3Si_5$, thus firmly confirming that these two structures are fundamentally distinct crystallographic phases.

To further verify that the *P-62c*-$U_3Si_5$ structure identified in this study represents a novel crystal configuration distinct from the $AlB_2$-type, we utilized the XRD simulation function of the VESTA software [56] to perform simulated analyses of the diffraction patterns for both structures. The results are shown in Fig. 3(g): The relative positions of the characteristic peaks in the simulated XRD patterns of the two $U_3Si_5$ structures are largely consistent with the experimentally measured results [57]. Moreover, previous experimental reports have confirmed the existence of distorted 12-membered silicon-ring structures in lanthanide silicides such as $Er_3Si_5$ and $Yb_3Si_5$ [58,59]. This feature is consistent with the new structure predicted in this study, providing additional evidence for the plausibility of this structural model. In addition, our study employed ab initio molecular dynamics (AIMD) simulations to validate the thermal stability of the distorted 12-silicon ring at 500 K (Fig. 3(h)). By comparing the structural evolution of the distorted 12-silicon ring after removal of the uranium atoms, we confirmed that the predicted *P-62c*-$U_3Si_5$ structure exhibits excellent high-temperature structural stability.

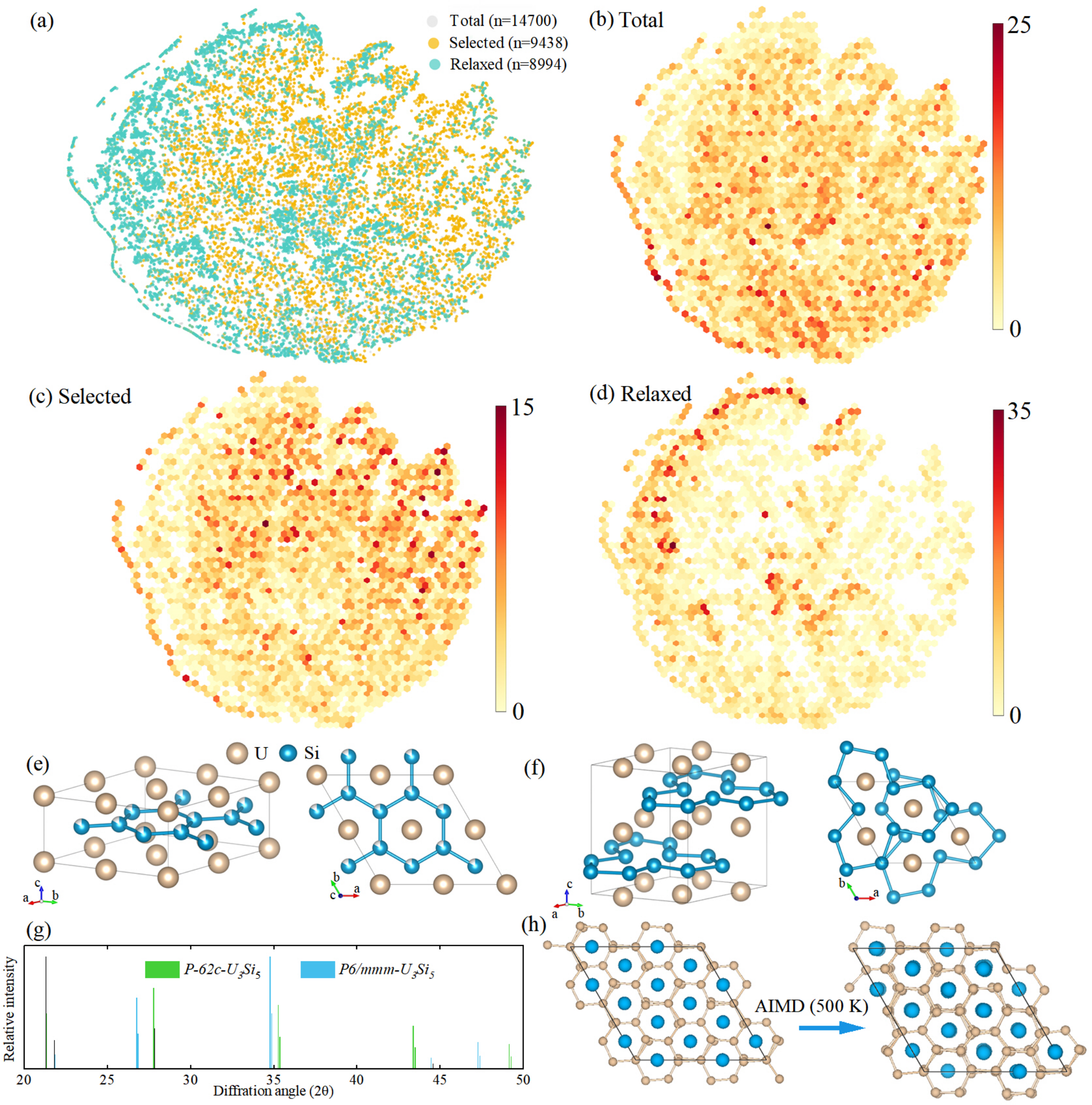

**Fig 3.** (a) t-SNE visualization comparing the distributions of the total generated, selected, and relaxed structures. Density distributions for (b) the total generated structures, (c) the selected structures, and (d) the relaxed structures. (e) The Schematic diagrams of $U_3Si_5$ with $USi_2$-type. (f) The crystal structure of *P*-62*c*-$U_3Si_5$ predicted in this work. (g) Simulated XRD patterns of $U_3Si_5$ with $USi_2$-type and *P*-62*c* space group. (h) The AIMD simulation results of *P*-62*c*-$U_3Si_5$.

Table 1 Lattice parameters of the *P-62c*-$U_3Si_5$ and $AlB_2$-type $U_3Si_5$

| | a (Å) | c (Å) | Energy eV/atom |
|---|---|---|---|
| *P-62c*-$U_3Si_5$ (gewum) | 3.665 | 4.258 | -7.841 |
| $AlB_2$-type $U_3Si_5$ （PBE） | 3.783 | 3.947 | -7.738 |
| *P-62c*-$U_3Si_5$ (gewum) (PBE +$U_{eff}$ 4 ev) | 3.734 | 4.169 | -6.422 |
| $AlB_2$-type $U_3Si_5$ (PBE +$U_{eff}$ 4 ev) | 3.842 | 3.969 | -6.434 |
| Experiment-1 | 3.847 | 4.072 | [57] |
| Experiment-2 | 3.878 | 4.031 | [57] |

### 3.3 Case 3：Crystal Structure Prediction of $ThH_{10}$ at 150 GPa via SRSS

To illustrate the capability of GEWUM for fixed-composition structure prediction in high-pressure hydride systems, we consider three-dimensional $ThH_{10}$ [60] at 150 GPa as a representative test case. This example is employed to assess the performance of the code in large-scale structure generation, rapid screening, and stability identification under extreme-pressure conditions. Unlike case 1, here we use the SRSS method to significantly reduce the number of structures in optimization procedure. The detailed process of the SRSS method can be found in [34].

Step 1: Candidate Structure Generation

For each space group, 200 trial structures were generated, yielding a total of 45,800 attempts. From these, 29,400 valid initial structures were ultimately obtained. Remarkably, this entire procedure required only ~30 s on a single 64-core CPU node.

Step 2: Representative-Structure Extraction and Compression of the Search Space

To enhance screening efficiency and reduce the computational burden, GEWUM extracts structural feature descriptors and combines them with a machine-learning clustering algorithm to identify representative configurations, thereby balancing the breadth of the search against the cost of subsequent optimization. In contrast to the Al-Sc-N system using the K-means strategy, HDBSCAN is considerably in $ThH_{10}$ case. [34] In the present case, representative structures and noisy structures were retained for each space group, compressing the 26,400 valid initial structures into a reduced set of 5,280. This step required only ~1 min. The t-SNE visualization, shown in Fig. 4(a), reveals that the screened structures remain uniformly distributed across the feature space, indicating that the diversity of the original structural pool is well preserved. Use the following command:

*gewum RD –mode select*

Step3: Structural Optimization at High-Pressure and Screening of Low-Enthalpy Candidates

GEWUM carried out high-pressure structural optimization for the retained 5,280 candidate structures at 150 GPa. This step required approximately 4 h on a single 64-core CPU node. Upon completion of the optimization, the program automatically compiled the converged structures, ranked them according to their enthalpies, and extracted the 10 lowest-enthalpy candidates; this post-processing stage required only ~28 s. The resulting low-enthalpy set spans a variety of symmetry classes, including $F$-43$m$, $P4_22_12$, $I4/mmm$, $Fm$-3$m$, and $P63/mmc$.

Step 4: Assessment of Dynamical Stability

We further performed phonon-spectrum calculations for the 10 selected representative structures, with the total phonon calculation requiring approximately only 4 min at one CPU node. The results reveal that structures such as $F$-43$m$, $P4_22_12$, $I4/mmm$, and $P6_3/mmc$ exhibit imaginary phonon modes to varying degrees, indicating that they are dynamically unstable at 150 GPa. By contrast, the $Fm$-3$m$-$ThH_{10}$ phase reported in the literature displays substantially superior dynamical stability, as illustrated in Figs. 4(b-d).

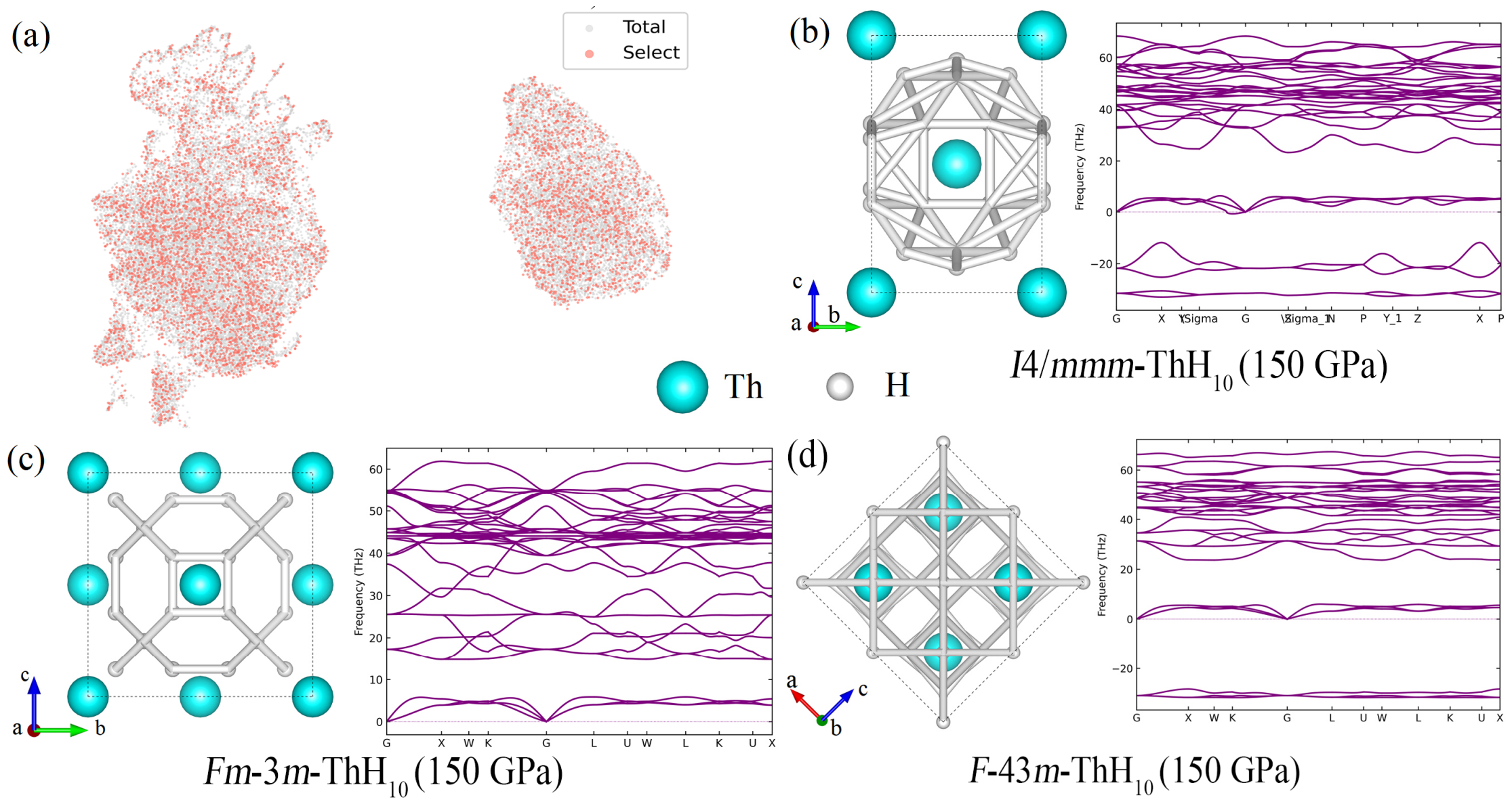

**Fig. 4.** (a) UMAP visualization for feature analysis and representative-structure extraction from the candidate pool. (b-d) Schematic illustration of some dynamically unstable and dynamically stable structures.

## Case 4: Validation of the Thermophysical-Property Module

To assess the applicability of GEWUM to thermophysical-property calculations, we benchmark two representative modules, namely, TC module and QHA module. By replacing computationally demanding force-constant evaluations with machine-learning interatomic potentials, combined with approximate schemes such as the relaxation-time approximation (RTA), GEWUM enables rapid evaluation of thermal transport properties and thereby establishes a thermophysical workflow suitable for high-throughput screening.

Taking cubic SiC as an example, the calculated lattice thermal conductivity decreases monotonically with increasing temperature over the range 300-1700 K, in accord with the overall experimental trend [61]. This result indicates that the module is capable of reasonably capturing the suppression of thermal conductivity induced by enhanced phonon scattering at elevated temperatures, as shown in Fig. 5(a). In addition, a comparison of results obtained with different machine-learning potentials for $U_3Si_2$ shows that MatterSim-1M [62], MatterSim-5M [62], and DPA3-OMat24 [63] yield broadly consistent temperature-dependent trends, while exhibiting noticeable discrepancies in absolute magnitude and the degree of anisotropy.

This finding suggests that lattice thermal conductivity is highly sensitive to the accuracy of the underlying machine-learning potential, as illustrated in Fig. 5(b). Beyond thermal conductivity, GEWUM also implements modules for calculating the thermal expansion coefficient and Gibbs free energy on the basis of machine-learning potentials. Using $U_3Si_2$ as an example (Fig. 5(c)), the volumetric thermal expansion coefficient is found to increase monotonically with temperature and to remain in overall agreement with reference experimental data in both order of magnitude and temperature dependence, indicating that this module can adequately characterize the high-temperature thermodynamic response of materials. The typical computational cost of the thermophysical module is approximately 3-30 min/structure on a single CPU node, substantially lower than that of conventional first-principles thermal-transport calculations. GEWUM thus not only supports structure search and stability assessment, but can also be extended to the automated evaluation of thermophysical properties, providing a continuous computational framework for materials discovery from candidate structures to functional properties.

It is important to emphasize that GEWUM is not intended to replace existing, well-established automated materials simulation platforms. Rather, in the context of the rapid advancement of machine learning potentials, it aims to serve as an accessible auxiliary platform for computational materials science, particularly for newcomers to the field. Beyond integrating the SRSS crystal structure prediction workflow, GEWUM offers a suite of ASE-based utilities that enable users to readily employ various uMLIPs for tasks ranging from crystal structure prediction to the calculation and analysis of material properties such as thermal conductivity. The design of GEWUM prioritizes workflow usability, featuring simplified parameter configuration, seamless integration between computational stages, and built-in post-processing and visualization tools. These functionalities are not meant to supersede specialized simulation software, but rather to lower the barrier to entry for applying MLPs in practical materials research. This allows users without extensive experience in constructing complex workflows to efficiently complete the essential pipeline-from structure generation and computation to screening, analysis, and visualization.

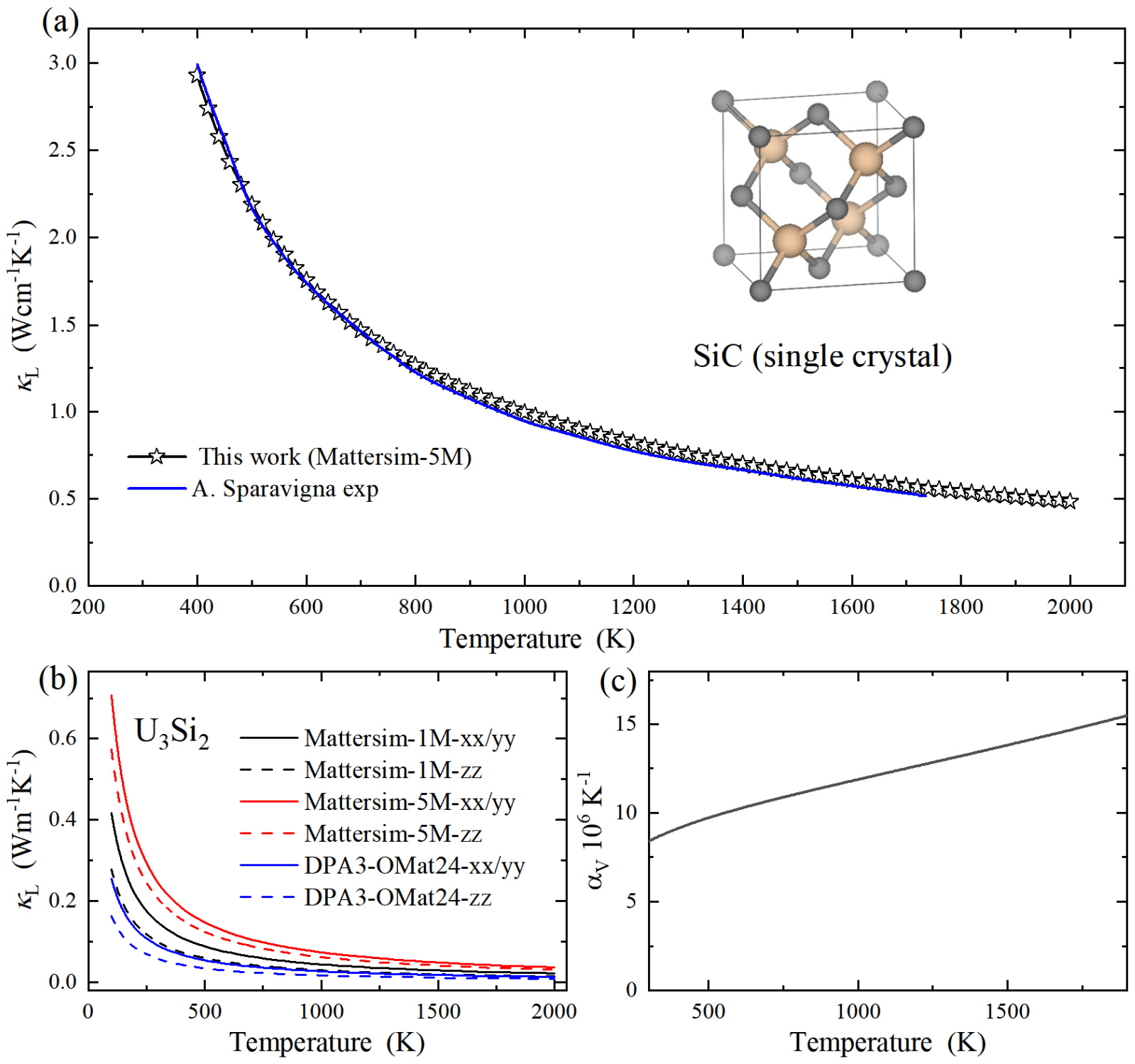

**Fig. 5.**Validation of thermophysical property calculations using GEWUM. (a) Lattice thermal

conductivity of cubic SiC, (b) Comparison of anisotropic thermal conductivity in $U_3Si_2$ predicted by different uMLIPs (MatterSim-1M, MatterSim-5M, and DPA3-OMat24). (c) Volumetric thermal expansion coefficient of $U_3Si_2$ demonstrating good agreement with experimental trends.

# CONCLUSION

In summary, we have developed an open-source platform GEWUM. By integrating uMLIPs with the SRSS methodology, GEWUM provides a robust, end-to-end solution for automated structure prediction and property characterization. Beyond computation, the platform embeds a suite of integrated visualization tools, such as Sankey diagrams for space-group evolution tracking, violin plots for energy distribution comparison, and t-SNE/UMAP projections for structural diversity monitoring that allow users to directly inspect and interpret the statistical characteristics of large-scale screening campaigns without external post-processing. We demonstrated GEWUM's versatility through four case studies: the discovery of polymorphs in the ternary Al-Sc-N system, the identification of a *P*-62*c* phase of $U_3Si_5$, the high-pressure prediction of $ThH_{10}$, and validation of lattice thermal conductivity and thermal expansion coefficient calculations. By lowering the barrier to large-scale computational screening, GEWUM offers a pathway to accelerating crystal structure prediction and the discovery of functional materials for energy and sustainability applications.

# Code Availability

GEWUM is an open-source Python package distributed under the MIT License and is available at https://github.com/JesseOOPP/GEWUM. The software requires Python $\geq$ 3.8 and depends on a set of well-established scientific computing libraries, including NumPy, SciPy, pandas, matplotlib, ASE, pymatgen, PyXtal, spglib, Phonopy, Phono3py, scikit-learn, and the Materials Project API client (mp-api). The package can be installed in editable mode directly from the source repository:

```
git clone https://github.com/JesseOOPP/GEWUM
cd gewum
pip install -e .
```

Machine learning interatomic potential backends, such as DPA3, SevenNet, MatterSim, MACE or UPET are treated as optional dependencies and should be installed according to the user's hardware environment and preference. GEWUM is designed to be backend-agnostic with respect to uMLIPs, requiring only ASE Calculator interface compatibility. The current stable release is version 1.0.0.

# Program summary

**Program Title:** GEWUM
**Licensing provisions:** MIT
**Programming language:** Python 3
**External routines/libraries:** ASE, pymatgen, pyxtal, spglib, phonopy, phono3py, SciPy, NumPy, scikit-learn, HDBSCAN, Dscribe, mp-api, universal machine learning potential (uMLIP, optional for crystal relaxation).
**Nature of problem:** Automated crystal structure prediction and thermophysical property validation using uMLIPs.
**Solution method:** GEWUM implements a unified workflow platform that integrates the Selective Random Structure Search (SRSS) strategy with uMLIPs for efficient exploration of chemical spaces. The platform automates the complete pipeline: (1) random crystal structure generation across all 230 space groups; (2) diversity preserving structure selection using geometric descriptors and machine learning clustering; (3) structure relaxation driven by uMLIPs (e.g., MatterSim, DPA3 or MACE); (4) thermodynamic stability assessment through energy-above-hull ($E_{hull}$) calculations; (5) dynamic stability validation via phonon dispersion analysis; and (6) advanced property calculations including elastic constants, quasi-harmonic approximation (QHA), thermal conductivity (via third-order force constants), and molecular dynamics simulations. All workflows are orchestrated through a modular command-line interface with native SLURM job scheduling support.
**Restrictions:** Requires a SLURM-based HPC cluster environment for full workflow execution. The uMLIP-driven relaxation and property calculations require the optional uMLIP backbone like MatterSim package.
**Unusual features:** GEWUM provides an end-to-end automated workflow from random structure generation to property validation, eliminating the need for manual intervention between computational stages. The platform features native SLURM integration with automatic job script generation and dependency management. It supports both high-throughput screening and targeted property calculations. A modular architecture allows users to execute individual workflow stages independently or chain them into custom pipelines through eight CLI commands (cifgen, RD, PT, ELA, QHA, TC, MD, FT). The platform bridges uMLIPs with automated high-throughput workflows, enabling rapid materials exploration at near-DFT accuracy with significantly reduced computational cost.

## Conflicts of interest

There are no conflicts of interest to declare.

## Acknowledgements

This work was supported in part by the National Natural Science Foundation of China (52271037), the Space Application System China Manned Space Program (ZDBS-ZRKJZ-TLC021), the Shaanxi Provincial Natural Science Fundamental Research Program, China (2025SYS-SYSZD-098), the Basic Research Program of Jiangsu (Grant No. BK20240395), the Jiangsu Funding Program for Excellent Postdoctoral Talent (Grant No. 2025ZB701) and Opening Grant of Zhejiang Key Laboratory of Data-Driven High-Safety Energy Materials and Applications (OG2024008). Calculations were performed on Sugon HPC clusters equipped with HYGON X86 32-core processors (2.5 GHz) and at the Beijing Super Cloud Computing Center.